\newif\ifemulateapj
\emulateapjtrue        

\ifemulateapj
    \documentclass[iop]{emulateapj}
\else
    \documentclass[12pt,preprint]{aastex}
\fi
\usepackage[hyperfootnotes=false,naturalnames=true,pdfstartview=FitH,pdfpagemode=UseNone]{hyperref}

\newcommand{\peras}{\ensuremath{\mathrm{arcsec}^{-1}}}
\newcommand{\kms}{\mbox{km\ s$^{-1}$}}
\newcommand{\kmsMpc}{\kms~\mbox{Mpc}$^{-1}$}
\newcommand{\Msun}{\ensuremath{M_{\odot}}}
\newcommand{\Msunyr}{\Msun~\mbox{yr}$^{-1}$}
\newcommand{\Lstar}{\ensuremath{L_{*}}}
\newcommand{\multasec}[2]{\mbox{#1\arcsec$\,\times\,$#2\arcsec}}
\newcommand{\multamin}[2]{\mbox{#1\arcmin$\,\times\,$#2\arcmin}}
\newcommand{\dg}{\ensuremath{^\circ}}
\newcommand{\HST}{\textit{HST}}
\newcommand{\Spitzer}{\textit{Spitzer}}

\newcommand{\Bband}{\ensuremath{B_{435}}}
\newcommand{\gband}{\ensuremath{g_{475}}}
\newcommand{\Vband}{\ensuremath{V_{555}}}
\newcommand{\rband}{\ensuremath{r_{625}}}
\newcommand{\iband}{\ensuremath{i_{775}}}
\newcommand{\zband}{\ensuremath{z_{850}}}
\newcommand{\Jband}{\ensuremath{J_{125}}}
\newcommand{\Hband}{\ensuremath{H_{160}}}
\newcommand{\iraci}{3.6~\micron}
\newcommand{\iracii}{4.5~\micron}

\newcommand{\zmJ}{\ensuremath{\zband - \Jband}}
\newcommand{\JmH}{\ensuremath{\Jband - \Hband}}

\def\figa {
   \begin{figure*}[ht]
   \epsscale{1.0}
   \plotone{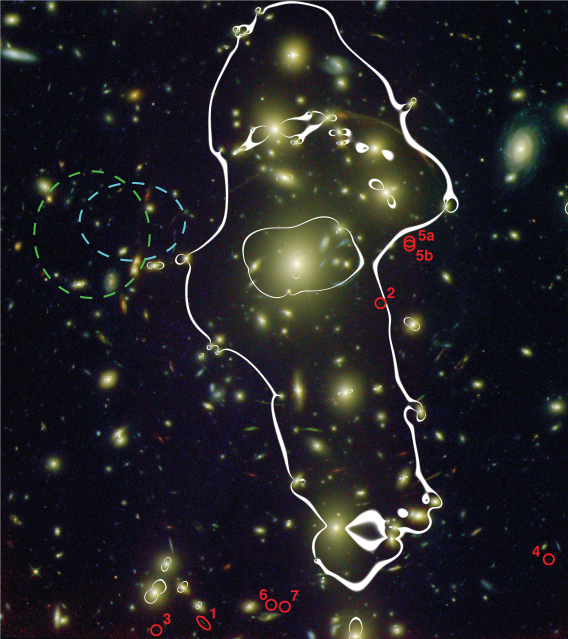}
   \caption{Color image (\zband \Jband \Hband) of the galaxy cluster Abell~1703 ($z=0.28$).  The locations of the high-redshift \zband-dropout candidate galaxies are marked by red circles and ellipses.  The image field of view is \multasec{123}{136} and is shown with P.A.$=152\dg$.  The white contours represent the critical curves at $z\sim7$.  The dashed cyan and green ellipses denote the regions where the strong lensing model of \cite{Zitrin2010} predicts counterimages for A1703-zD2 and A1703-zD5a/5b, respectively.}
   \label{fig:a1703}
   \end{figure*}
}

\def\figb {
   \begin{figure*}[ht]
   \epsscale{1.0}
   \plotone{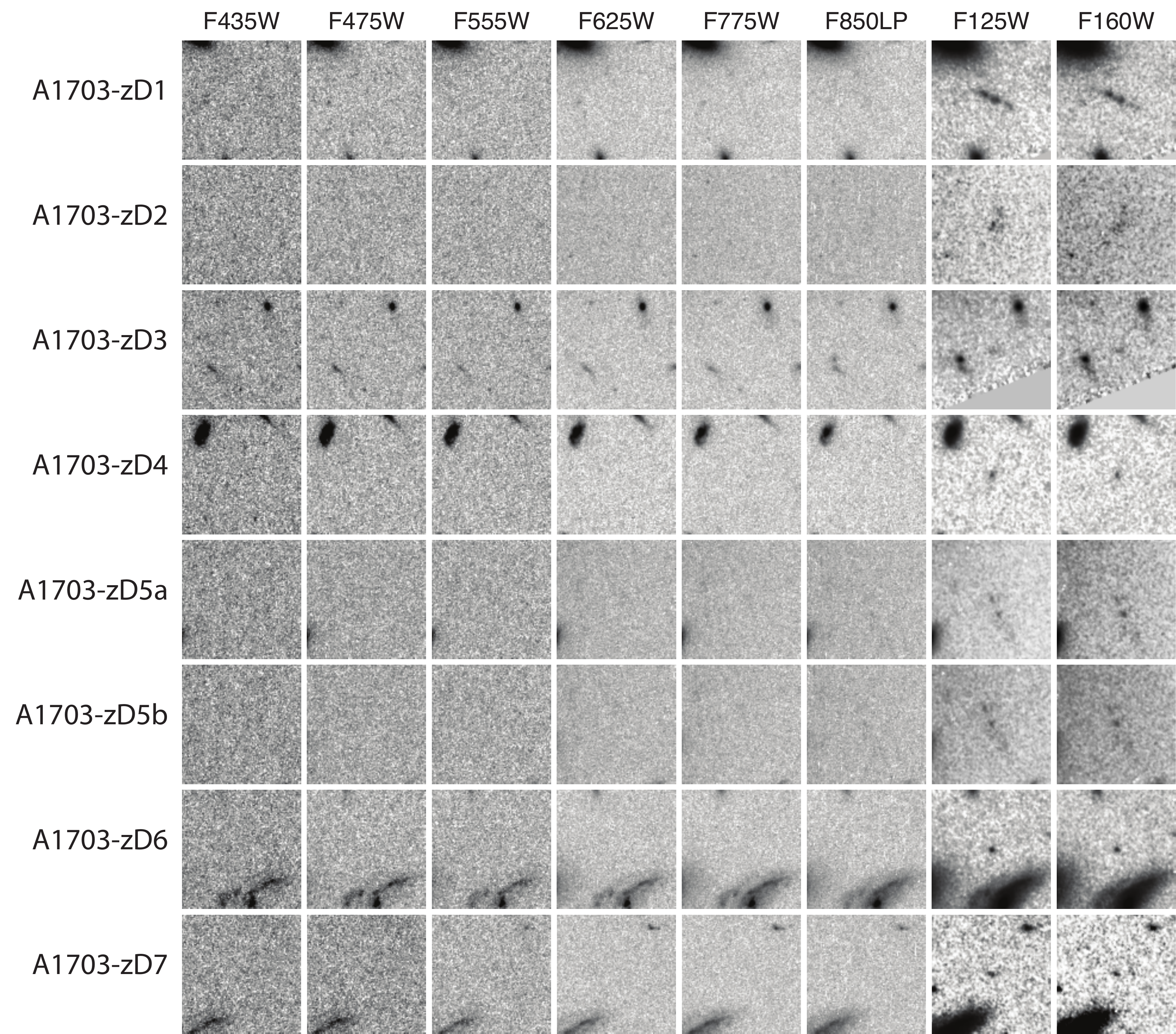}
   \caption{Postage stamp cutout images of the high-redshift \zband-dropout candidate galaxies from the \HST\ ACS and WFC3/IR data.  The cutout images are \multasec{6}{6}, corresponding to 31.4~kpc on a a side at $z=7$, and are shown with P.A.$=130\dg$.  As discussed in the text, A1703-zD5a/5b most likely represent two star-forming knots within a single source.}
   \label{fig:stamps}
   \end{figure*}
}

\def\figc {
   \begin{figure}[ht]
   \epsscale{1.0}
   \plotone{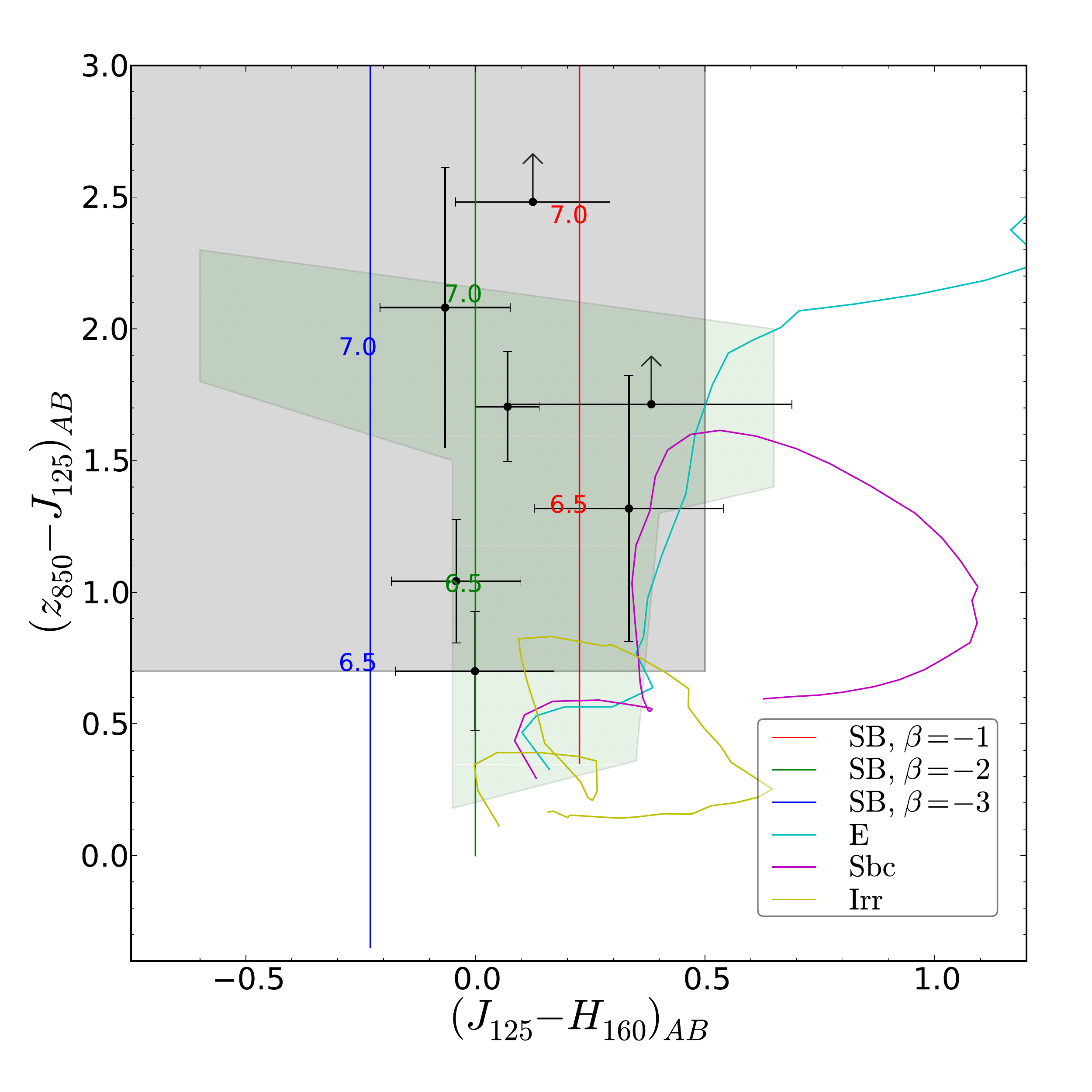}
   \caption{\zmJ\ vs. \JmH\ two-color diagram used to select the \zband-band dropout candidates.  The error bars and lower limits are $1\sigma$.  The gray region represents the \zmJ\ and \JmH\ colors of the selection criteria.  The blue, green, and red lines show the expected colors of star-forming galaxies with UV-continuum slopes of $\beta$ = -1, -2, and -3 ($f_{\lambda} \propto \lambda^{\beta}$), respectively.  The cyan, magenta, and yellow lines depict the expected colors
of low-redshift interloper galaxies.  The green region indicates the colors of low-mass L,T dwarf stars \citep[e.g.,][]{Knapp2004}.  While there is considerable overlap in the colors of L,T stars, only one of the \zband-dropout candidates (A1703-zD6) is unresolved.  However, this candidate has been subsequently spectroscopically confirmed at $z=7.045$ \cite{Schenker2011}.}
   \label{fig:colcol}
   \end{figure}
}

\def\figd {
   \begin{figure}[ht]
   \epsscale{1.0}
   \plotone{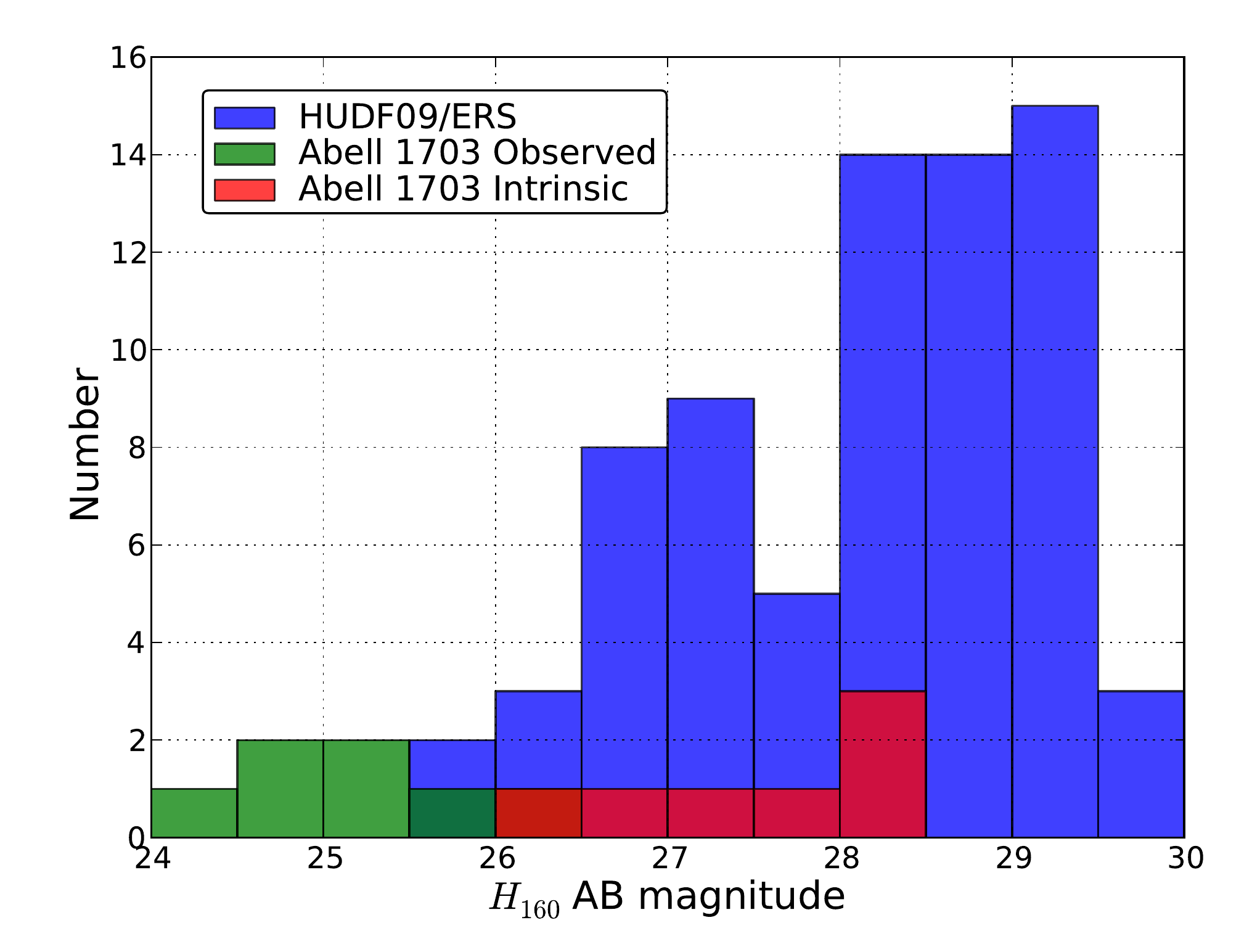}
   \caption{Histogram of the observed and intrinsic (unlensed) \Hband\ magnitudes for the seven \zband-dropout ($z\sim7$) candidate galaxies identified behind Abell~1703 (green and red) compared with the 73 $z\sim7$ candidates found in the HUDF09 and its two parallel fields and the WFC3/IR Early Release Science observations (blue) \citep{Bouwens2011a}.}
   \label{fig:hist}
   \end{figure}
}

\def\fige {
   \begin{figure}[ht]
   \epsscale{0.9}
   \plotone{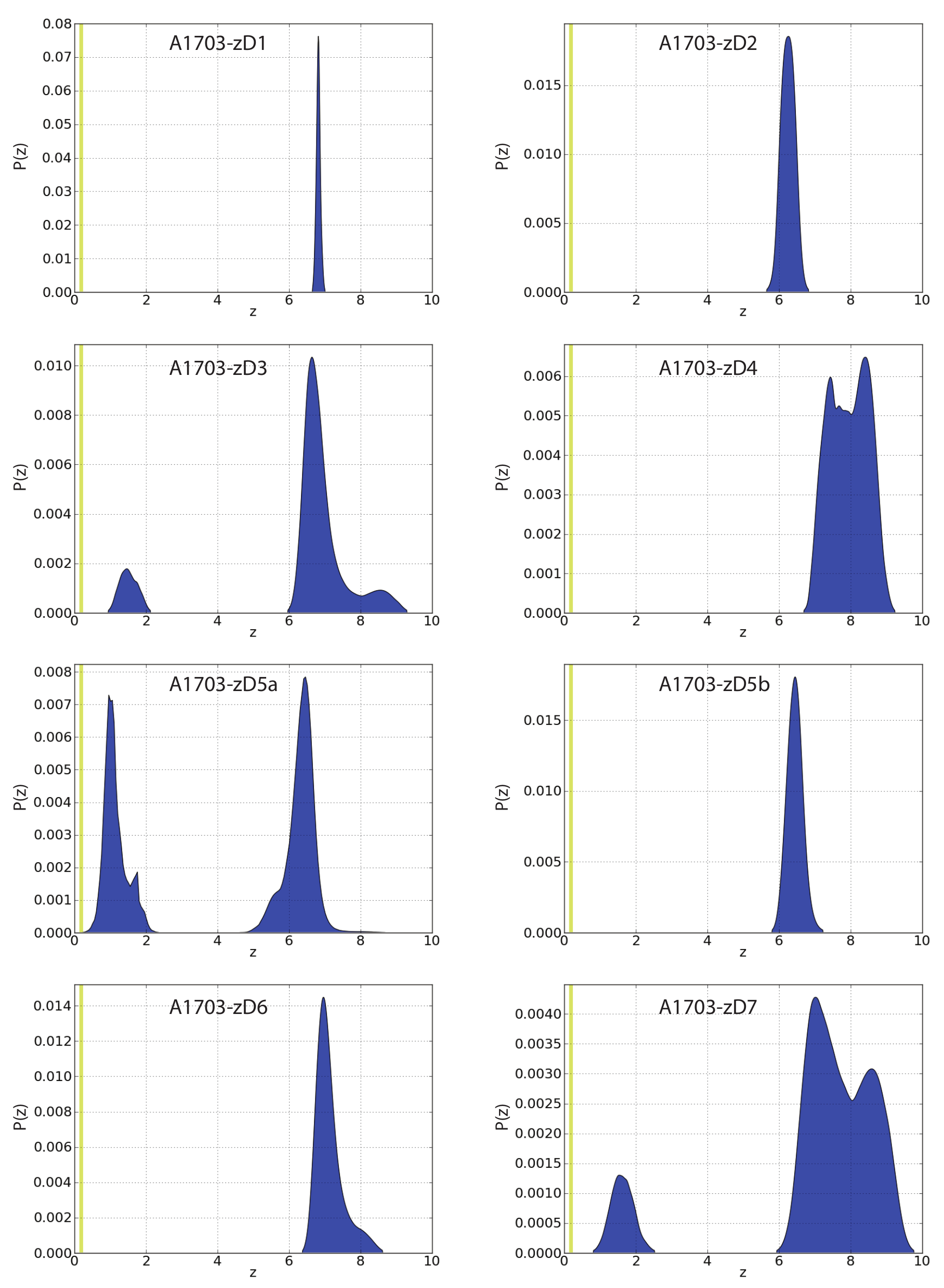}
   \caption{Probability distributions of the photometric redshifts for each of the candidates.  The vertical yellow line represents the redshift of the Abell~1703 cluster ($z=0.28$).  As discussed in the text, A1703-zD5a/5b most likely represent two star-forming knots within a single source.  The A1703-zD5a component shows the highest probability of being at low redshift due to its modest $2.1\sigma$ detection in the F555W band, which we attribute to a statistical fluctuation as described in the text.}
   \label{fig:pz}
   \end{figure}
}

\def\figf {
   \begin{figure}[ht]
   \epsscale{1.0}
   \plotone{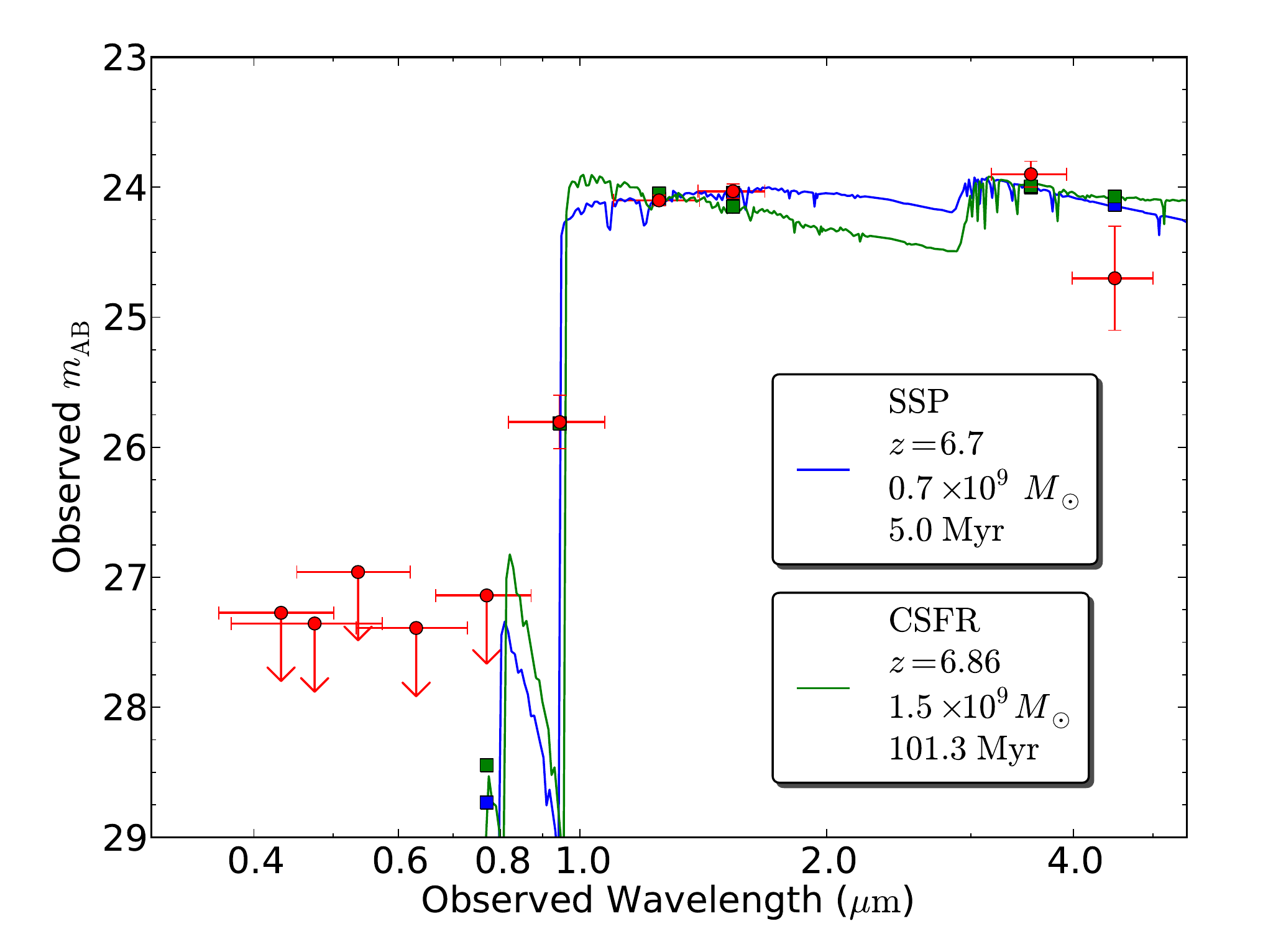}
   \caption{Best-fit stellar population models to the observed multiband photometry of A1703-zD1.  The models assume a \cite{Salpeter1955} IMF with subsolar ($Z = 0.2Z_{\sun}$) metallicity.  The $2\sigma$ upper limits are shown for the ACS optical non-detections, but the models were fit to the measured fluxes and errors.  The stellar masses are intrinsic values, corrected for the cluster magnification.  The larger $\chi^{2}_{\nu}$ values ($1.3-2.4$) for these models results from the poor fit to the IRAC 4.5\micron\ band data.}
   \label{fig:zd1}
   \end{figure}
}

\def\figg {
   \begin{figure}[ht]
   \epsscale{1.0}
   \plotone{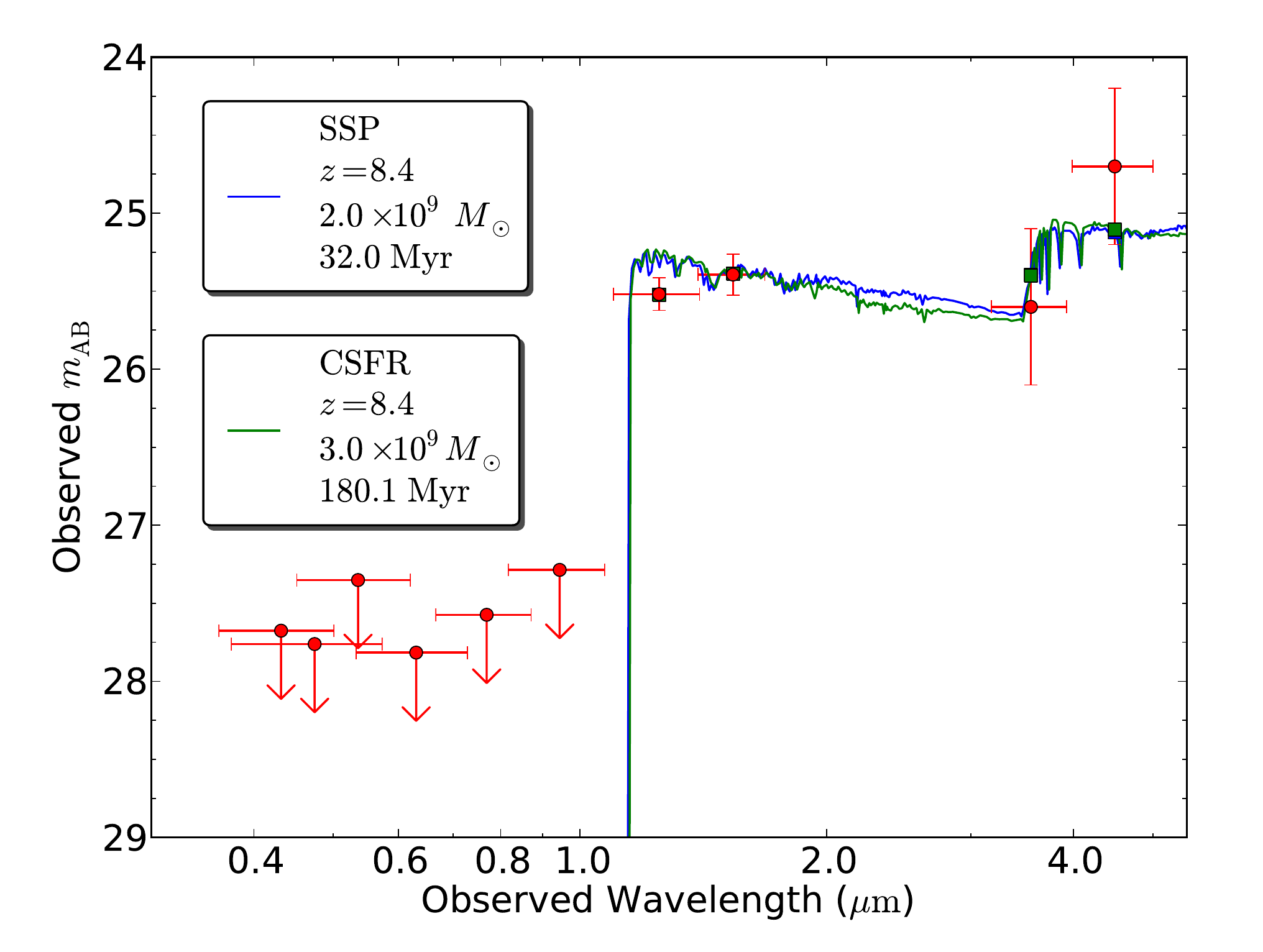}
   \caption{Best-fit stellar population models to the observed multiband photometry of A1703-zD4.  The models assume a \cite{Salpeter1955} IMF with subsolar ($Z = 0.2Z_{\sun}$) metallicity.  The $2\sigma$ upper limits are shown for the ACS optical non-detections, but the models were fit to the measured fluxes and errors.  The stellar masses are intrinsic values, corrected for the cluster magnification.}
   \label{fig:zd4}
   \end{figure}
}

\def\figh {
   \begin{figure}[ht]
   \epsscale{1.0}
   \plotone{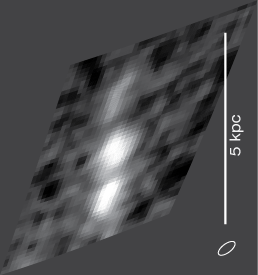}
   \caption{Source-plane reconstruction of A1703-zD1 in the WFC3/IR \Jband\ band.  The source has an extended morphology that spans $\sim0.74\arcsec$ ($\sim4$~kpc) in the source plane at $z=6.7$.  A1703-zD1 is comprised of three separate resolved knots, each with radius $r\sim0.08\arcsec$ ($\sim0.4$~kpc).  Of course the physical size and structure we infer for this candidate is somewhat dependent on the details of the gravitational lensing model.  The white ellipse denotes the WFC3/IR \Jband\ PSF in the source plane ($0.16\arcsec$ FWHM in the image plane).}
   \label{fig:zd1src}
   \end{figure}
}

\def\taba {
\ifemulateapj
   \tabletypesize{\scriptsize}
   \begin{deluxetable*}{lccccccccc}
\else
   \begin{deluxetable}{lccccccccc}
   \rotate
   \tabletypesize{\scriptsize}
\fi
\tablecolumns{10}
\tablewidth{0pt}
\tablecaption{Observed Photometry of High-Redshift Candidates}
\tablehead{\colhead{Candidate} & \colhead{R.A.} & \colhead{Dec} & \colhead{\zband} & \colhead{\Jband} & \colhead{\Hband} & \colhead{\iraci} & \colhead{\iracii} & \colhead{$\mu$\tablenotemark{a}} & \colhead{$z_{phot}$\tablenotemark{b}}}
\startdata
A1703-zD1                   & 13:14:59.4183 & 51:50:00.843 & $25.8 \pm  0.20$ & $24.1 \pm 0.04$ & $24.0\pm 0.06$ & $23.9\pm 0.1$ & $24.7\pm 0.4$ & $9.0^{+0.9}_{-4.4}$    & $6.7^{+0.2}_{-0.1}$ \\
A1703-zD2                   & 13:15:06.5089 & 51:49:17.960 & $25.6 \pm  0.20$ & $24.9 \pm 0.10$ & $24.9\pm 0.14$ & \nodata       & \nodata       & $24.8^{+43.6}_{-10.4}$  & $6.4^{+0.1}_{-0.3}$ \\
A1703-zD3                   & 13:14:58.3860 & 51:49:57.740 & $26.8 \pm  0.48$ & $25.5 \pm 0.14$ & $25.1\pm 0.15$ & \nodata       & \nodata       & $ 7.3^{+1.3}_{-2.5}$   & $6.7^{+1.6}_{-0.2}$ \\
A1703-zD4                   & 13:15:07.1889 & 51:50:23.552 & $> 28.0$         & $25.5 \pm 0.10$ & $25.4\pm 0.13$ & $25.6\pm 0.5$ & $24.7\pm 0.5$ & $ 3.1^{+0.2}_{-1.0}$ & $8.4^{+0.9}_{-1.4}$ \\
A1703-zD5a\tablenotemark{c,d} & 13:15:07.7650 & 51:49:09.333 & $26.7 \pm  0.34$ & $25.6 \pm 0.12$ & $25.7\pm 0.18$ & \nodata       & \nodata       & $39.0^{+26.9}_{-15.0}$  & $6.5^{+0.2}_{-0.3}$ \\
A1703-zD5b\tablenotemark{c}   & 13:15:07.7036 & 51:49:10.139 & $26.3 \pm  0.29$ & $25.3 \pm 0.11$ & $25.3\pm 0.15$ & \nodata       & \nodata       & $26.9^{+19.9}_{-8.0}$   & $6.5^{+0.2}_{-0.2}$ \\
A1703-zD6\tablenotemark{e}  & 13:15:01.0068 & 51:50:04.353 & $27.9 \pm  0.53$ & $25.8 \pm 0.08$ & $25.9\pm 0.12$ & \nodata       & \nodata       & $ 5.2^{+0.3}_{-0.9}$   & $7.0^{+0.6}_{-0.2}$ \\
A1703-zD7                   & 13:15:01.2696 & 51:50:06.052 & $> 28.5$         & $26.8 \pm 0.22$ & $26.4\pm 0.21$ & \nodata       & \nodata       & $ 5.1^{+0.4}_{-0.9}$   & $8.8^{+0.2}_{-1.7}$
\enddata
\tablecomments{The sources without quoted IRAC magnitudes either suffer from significant confusion from neighboring sources or do not show an especially prominent ($>2\sigma$) detection.}
\tablenotetext{a}{The magnification errors represent the extreme values obtained from the minimum and maximum magnifications obtained within $\pm 0.5\arcsec$ of each candidate and assuming $\Delta z \pm 1.0$ for the source redshifts.}
\tablenotetext{b}{Photometric redshifts determined from the BPZ code \citep{Benitez2000}.  Because of the limited depth of the optical data, there is a small chance that some of the sources could be at low redshift (see \S~\ref{sec:bpz}).}
\tablenotetext{c}{As discussed in the text, A1703-zD5a/5b most likely represent two star-forming knots within a single source.  The magnification of the combined A1703-zD5 source is $31.9^{+20.3}_{-10.6}$.}
\tablenotetext{d}{This component of the zD5 candidate is detected in the \Vband\ band at low significance ($2.1\sigma$), but not in the overlapping \gband\ and \rband\ bands.  Thus, the slight \Vband-band detection is likely a statistical fluctuation.}
\tablenotetext{e}{This candidate is spectroscopically confirmed to be at $z = 7.045$ \cite{Schenker2011}.}
\label{tbl:photometry}
\ifemulateapj
    \end{deluxetable*}
\else
    \end{deluxetable}
\fi
}

\def\tabb {
\ifemulateapj
    \tabletypesize{\footnotesize}
    \begin{deluxetable*}{lccccccc}
\else
    \begin{deluxetable}{lccccccc}
\fi
\tablecolumns{8}
\tablewidth{0pt}
\tablecaption{Best-fit Stellar Population Model Results}
\tablehead{ & & & Mass\tablenotemark{c} & Age$_{w}$\tablenotemark{d} & SFR & & \\
Candidate & SFH\tablenotemark{a} & $z_{phot}$\tablenotemark{b} & ($10^{9} \Msun$) & (Myr) & (\Msunyr) & $A_V$ & $\chi^{2}_{\nu}$
}
\startdata
 A1703-zD1 & SSP  & $6.7 \pm 0.1$   & $0.7 \pm 0.1$   &  5.0    & \nodata        & $0.7 \pm 0.1$        & 1.3 \\
           & CSFR & $6.86 \pm 0.1$  & $1.5 \pm 0.05$  &  101.3   & $7.3 \pm 0.3$  & $0.00 \pm 0.00$        & 2.4 \\[0.5em]
 A1703-zD4 & SSP  & $8.4 \pm 0.3$   & $2.0 \pm 1.4$   &  32.0   & \nodata        & $0.09^{+0.29}_{-0.09}$ & 0.8 \\
           & CSFR & $8.4 \pm 0.3$   & $3.0 \pm 0.4$   &  180.1  & $8.2 \pm 1.2$  & $0.0 \pm 0.0$        & 0.8
\enddata
\tablecomments{Models assume a \cite{Salpeter1955} IMF with mass cutoffs of 0.1 and 100 \Msun\ and subsolar ($Z = 0.2 Z_{\sun} = 0.004$) metallicities.}
\tablenotetext{a}{Star-formation history:  simple (single-burst) stellar population (SSP) or constant star-formation rate (CSFR) models.}
\tablenotetext{b}{Photometric redshifts derived from the stellar population fitting.  These redshifts are consistent with those presented in Table~\ref{tbl:photometry}.}
\tablenotetext{c}{Best-fit stellar mass, corrected by the magnification at the fitted $z_{phot}$ redshift.}
\tablenotetext{d}{SFR-weighted mean stellar age \citep[cf.][]{Forster2004}.  For a CSFR model, the weighted age is simply half of the time elapsed since the start of star formation.}
\label{tbl:models}
\ifemulateapj
    \end{deluxetable*}
\else
    \end{deluxetable}
\fi
}

\ifemulateapj
    \journalinfo{Accepted to The Astrophysical Journal}
    \slugcomment{Accepted to The Astrophysical Journal}
    \shorttitle{HIGHLY MAGNIFIED HIGH-REDSHIFT GALAXIES}
    \shortauthors{BRADLEY ET AL.}
\fi

\begin{document}

\title{Through the Looking Glass:  Bright, Highly Magnified Galaxy Candidates at \lowercase{$z\sim7}$ Behind Abell 1703 \altaffilmark{1}}

\author{L.D.~Bradley\altaffilmark{2}, R.J.~Bouwens\altaffilmark{3}, A.~Zitrin\altaffilmark{4}, R.~Smit\altaffilmark{3}, D.~Coe\altaffilmark{2}, H.C.~Ford\altaffilmark{5}, W.~Zheng\altaffilmark{5}, G.D.~Illingworth\altaffilmark{6}, N.~Ben\'{i}tez\altaffilmark{7}, T.J.~Broadhurst\altaffilmark{8,9}}

\altaffiltext{1}{Based on observations made with the NASA/ESA {\em
Hubble Space Telescope}, obtained at the Space Telescope Science
Institute, which is operated by the Association of Universities
for Research in Astronomy under NASA contract NAS5-26555.  Based on
observations made with the {\em Spitzer Space Telescope}, which is
operated by the Jet Propulsion Laboratory, California Institute of
Technology under NASA contract 1407.}
\altaffiltext{2}{Space Telescope Science Institute, 3700 San Martin Drive, Baltimore, MD 21218.}
\altaffiltext{3}{Leiden Observatory, Leiden University, Postbus 9513, 2300 RA Leiden, Netherlands.}
\altaffiltext{4}{School of Physics and Astronomy, Tel Aviv University, Tel Aviv 69978, Israel.}
\altaffiltext{5}{Department of Physics and Astronomy, Johns Hopkins University, 3400 North Charles Street, Baltimore, MD 21218.}
\altaffiltext{6}{UCO/Lick Observatory, Department of Astronomy and Astrophysics, University of California Santa Cruz, Santa Cruz, CA 95064.}
\altaffiltext{7}{Instituto de Astrof\'{i}sica de Andaluc\'{i}a (CSIC), C/Camino Bajo de Hu\'{e}tor 24, Granada 18008, Spain.}
\altaffiltext{8}{Department of Theoretical Physics, University of Basque Country UPV/EHU, Leioa, Spain.}
\altaffiltext{9}{Ikerbasque, Basque Foundation for Science, 48011 Bilbao, Spain.}

\begin{abstract}
We report the discovery of seven strongly lensed Lyman break galaxy
(LBG) candidates at $z\sim7$ detected in {\em Hubble Space Telescope
(HST)} Wide Field Camera 3 (WFC3) imaging of Abell~1703.  The
brightest candidate, called A1703-zD1, has an observed (lensed)
magnitude of 24.0 AB ($26 \sigma$) in the WFC3/IR F160W band, making
it 0.2 magnitudes brighter than the \zband-dropout candidate recently
reported behind the Bullet Cluster and 0.7 magnitudes brighter than the
previously brightest known $z\sim7.6$ galaxy, A1689-zD1.  With a
cluster magnification of $\sim9$, this source has an intrinsic
magnitude of $\Hband = 26.4$~AB, a strong $\zband - \Jband$ break of
1.7 magnitudes, and a photometric redshift of $z\sim6.7$.
Additionally, we find six other bright LBG candidates with \Hband\
band magnitudes of $24.9-26.4$, photometric redshifts $z\sim6.4 -
8.8$, and magnifications $\mu\sim3-40$.  Stellar population fits to
the ACS, WFC3/IR, and \Spitzer/IRAC data for A1703-zD1 and A1703-zD4
yield stellar masses $(0.7 - 3.0) \times 10^{9}\ M_{\sun}$, stellar
ages $5-180$~Myr, and star-formation rates $\sim 7.8\ M_{\sun}$
yr$^{-1}$, and low reddening with $A_V \le 0.7$.  The source-plane
reconstruction of the exceptionally bright candidate A1703-zD1
exhibits an extended structure, spanning $\sim 4$~kpc in the
$z\sim6.7$ source plane, and shows three resolved star-forming knots
of radius $r \sim 0.4$~kpc.
\end{abstract}

\keywords{galaxies: high-redshift --- gravitational lensing: strong ---
galaxies: clusters: individual: Abell 1703}

\section{Introduction}

The recently installed Wide Field Camera 3 (WFC3) aboard the {\em
Hubble Space Telescope} has led to a significant increase in the
sample of $z\ga 7$ galaxy candidates in the past year
\citep{Oesch2010a, Bouwens2010b, Bunker2010, McLure2010,
Finkelstein2010, Bouwens2011a, Trenti2011, Yan2011}.  Already, more
than 132 $z\sim7-8$ Lyman-break galaxy (LBG) candidates \citep[][see
also \citealt{Lorenzoni2010, McLure2011}]{Bouwens2011a} have been
found in ultradeep WFC3/IR observations of the Hubble Ultra-Deep Field
(HUDF) and nearby fields, with even a few candidates at $z\sim8.5$ and
one at $z\sim10$ \citep{Bouwens2011b}.  These observations provide our
first glimpse of galaxies during the reionization epoch, showing a
rapidly evolving galaxy luminosity function (LF) and a declining
star-formation rate with increasing redshift \citep{Bouwens2011a,
Bouwens2011b}.  Recent studies of these $z\ga7$ galaxies have also
looked at their structure and morphologies \citep{Oesch2010b},
rest-frame $UV$-continuum slopes \citep{Bouwens2010a, Finkelstein2010}
and star-formation rates and stellar masses \citep{Labbe2010a,
Labbe2010b, McLure2011}.

The ultradeep observations are complemented by shallower wide-field
WFC3/IR surveys for $z\ga7$ galaxies such as the Brightest of
Reionizing Galaxies (BoRG) \citep{Trenti2011} and Hubble Infrared Pure
Parallel Imaging Extragalactic Survey (HIPPIES) \citep{Yan2011}, which
so far have uncovered 4 bright ($25.5-26.7$) $z\ga7.5$ galaxies over
130 arcmin$^2$.  Over the next three years, the Cluster Lensing And
Supernova survey with Hubble (CLASH; \citealt{Postman2011}) and Cosmic
Assembly Near-infrared Deep Extragalactic Legacy Survey (CANDELS;
\citealt{Grogin2011, Koekemoer2011}) Multi-Cycle Treasury (MCT)
programs will further augment the sample of bright $z\ga7$ galaxy
candidates and our understanding of the high-redshift universe.  These
wide-field surveys are needed to characterize the bright end of the
galaxy LF, where bright $z\ga7$ galaxies are rare.  Placing tighter
constraints of the number density of bright sources also helps break
the degeneracy between the characteristic luminosity \Lstar\ and the
faint-end slope $\alpha$, a parameter whose value is crucial in
determining the contribution of galaxies to the reionization of the
universe.

The use of massive galaxy clusters as ``cosmic" gravitational
telescopes has uncovered some the brightest ($\la26.5$) $z\ga5$
high-redshift galaxies to date \citep{Franx1997, Frye2002, Kneib2004,
Egami2005, Bradley2008, Zheng2009, Bouwens2009a, Hall2011} and some of
the most distant galaxies known at the time of their discovery
\citep{Franx1997, Kneib2004, Bradley2008}.  Gravitational lensing by
massive galaxy clusters can amplify both the size and flux of
background sources considerably.  The increased spatial resolution
allows high-redshift galaxies to be studied in unprecedented detail,
providing clear views of their sizes and morphologies
\citep[e.g.,][]{Franx1997, Kneib2004, Bradley2008, Zheng2009,
Swinbank2009}.  This was clearly demonstrated with a very high
source-plane resolution of 50~pc recently achieved by
\cite{Zitrin2011} for the $z=4.92$ galaxy behind MS1358.  Likewise,
the increased brightness can place $z\ga7$ galaxies within reach of
ground-based spectroscopic follow-up observations.

Luminous high-redshift galaxies are extremely valuable because their
spectra can provide direct measurements of the early star-formation
rate via Ly$\alpha$ and H$\alpha$ emission \citep{Iye2006} and the
evolution of metallicity via metal emission and absorption lines
\citep{Dow-Hygelund2005}.  The spectra of $z\ga7$ objects also
pinpoint the epoch of the intergalactic medium (IGM) reionization
through the effect of neutral hydrogen in inhibiting the emission of
Ly$\alpha$ from galaxies \citep{Santos2004, Malhotra2004, Stark2010}.
A truly neutral IGM will produce a damped Ly$\alpha$ absorption
profile \citep{Miralda-Escude1998} that can be measured even at a low
spectral resolution.

Here we present the discovery of seven bright strongly lensed LBG
candidates at $z\sim7$ behind the massive galaxy cluster Abell~1703.
The brightest candidate, A1703-zD1, is observed at 24.0 AB in the
\Hband\ band, making it 0.2 magnitudes brighter than the
\zband-dropout candidate recently reported behind the Bullet Cluster
\citep{Hall2011} and 0.7 magnitudes brighter than the previously
brightest known $z\sim7.6$ galaxy A1689-zD1, found behind the massive
cluster Abell~1689 \citep{Bradley2008}.  This paper is organized as
follows.  We present the observations and photometry in \S~2 and
dropout selection in \S~3.  In \S~4 we discuss the source
magnifications.  We present the photometric redshifts in \S~5 and
stellar population synthesis models in \S~6.  The results and the
properties of the sources are discussed in \S~7.  We summarize our
results in in \S~8.  Throughout this work, we assume a cosmology with
$\Omega_{m} = 0.3$, $\Omega_{\Lambda} = 0.7$, and $H_{0} =
70$~\kmsMpc.  This provides an angular scale of 5.2~kpc
(proper)~\peras\ at $z = 7.0$.  All magnitudes are expressed in the AB
photometric system \citep{Oke1974}.

\ifemulateapj\figa\fi
\ifemulateapj\figb\fi
\ifemulateapj\figc\fi
\ifemulateapj\figd\fi
\ifemulateapj\taba\fi

\section{Observations and Photometry}

\subsection{\HST\ ACS and WFC3/IR Data}

We observed Abell~1703 ($z = 0.284$; \citealt{Allen1992}) with a
single field of the Advanced Camera for Surveys (ACS) Wide Field
Camera (WFC) in 2004 November as part of an ACS GTO program to study
five massive galaxy clusters (HST-GO10325).  The observations cover a
\multamin{3.4}{3.4} field of view and consist of 20 orbits divided
among six broadband filters: F435W (\Bband; 7050~s), F475W (\gband;
5564~s), F555W (\Vband ; 5564), F625W (\rband; 9834~s), F775W (\iband;
11128~s), and F850LP (\zband; 17800~s).  The ACS/WFC data were reduced
with our ACS GTO APSIS pipeline \citep{Blakeslee2003}.  The reductions
reach $5\sigma$ limiting magnitudes ($0.19\arcsec$ diameter aperture)
of 28.5, 28.6, 28.2, 28.6, 28.4, and 28.0 in the \Bband, \gband,
\Vband, \rband, \iband, and \zband\ bands, respectively.

We obtained WFC3/IR observations of Abell~1703 in the F125W (\Jband)
and F160W (\Hband) bands, each with an exposure time of 2812 s, in
2010 April with the primary goal to search for $z\sim7$ galaxies
(HST-GO11802).  The WFC3/IR observations cover the central
\multasec{123}{136} high-magnification region of the cluster (see
Fig.~\ref{fig:a1703}).  The depths of the WFC3/IR data reach $5\sigma$
limiting magnitudes of 27.3 and 26.9 in a $0.45\arcsec$ diameter
aperture for the \Jband\ and \Hband\ bands, respectively.

For the reduction of both the ACS and WFC3/IR data, we weight each
individual exposure by its inverse variance created from the sky
background, modulated by the flatfield variations, along with the read
noise and dark current.  The drizzle combination procedure uses these
inverse-variance images as weights and produces a final
inverse-variance image for the combined, drizzled image in each
filter.  These inverse-variance weight images are used by SExtractor
\citep{Bertin1996} for both source detection and photometry (see
\S~\ref{sec:photometry}).

\subsection{\Spitzer/IRAC Data}

We utilized archival \Spitzer/IRAC imaging of Abell~1703 (program
40311) obtained over three epochs between 2007 December and 2008 June.
We used the \Spitzer\ MOPEX calibration pipeline to combine the data
in the 3.6 and 4.5 \micron\ bands over the three epochs.  The total
exposure times were 18.9~ks in the 3.6 and 4.5 \micron\ bands,
reaching $5\sigma$ limiting magnitudes of 24.7 and 24.1, respectively.

\subsection{Photometry}
\label{sec:photometry}

We used SExtractor in dual-image mode for object detection and
photometry.  The detection image consisted of an inverse-variance
weighted combination of the WFC3/IR \Jband\ and \Hband\ images.  We
smoothed the ACS optical images to match the WFC3/IR images and
measured colors in small scalable Kron apertures (Kron factor of 1.2;
\citealt{Kron1980}).  We then correct the fluxes measured in these
smaller apertures to total magnitudes by using the flux measured in a
larger Kron aperture (factor of 2.5) on the detection image.
Likewise, we apply aperture corrections for light falling outside of
the large Kron aperture using the tabulated encircled energies
provided in the instrument handbooks.

We were able to obtain \Spitzer/IRAC fluxes for only two of the
brighter and isolated sources.  IRAC fluxes for these candidates were
obtained using the deblending algorithm of \cite{Labbe2006,
Labbe2010a}.  Briefly, this method involved using the higher
resolution \HST\ WFC3/IR images to create model IRAC images for the
source and its nearby neighbors (assuming no differences between the
structure or size of sources at 1.25 microns and IRAC wavelengths).
We then vary the normalization of each model image, for both the
source and its nearby neighbors, to fit the IRAC observations.
Finally, we subtract the best-fit model profiles for the neighbors and
perform photometry for the sources of interest in a
$2.5\arcsec$-diameter aperture.  The IRAC errors include a component
due to the modeling error.  The modeling error can be quite large for
sources near the edges of the WFC3/IR image because there is no source
template for IRAC sources found beyond the WFC3/IR field of view.

\section{Selection of \lowercase{$z\sim7$} \zband-band Dropout
Candidates}

We search for $z\sim7$ galaxies using a two-color \zband-dropout
selection criteria, based on the magnitudes measured in the small
scalable apertures.  We require candidates to have $\zmJ\ \ge 0.7$ and
$\JmH\ < 0.5$.  In addition, they must be undetected ($<2\sigma$) in
each optical ACS band, with not more than one band showing a
$>1.5\sigma$ detection.  Further, candidates must be detected at
$>5\sigma$ in the \Jband\ band.  In cases where an object is not
detected in a particular band, we assign the object with the $1\sigma$
detection limit to calculate object colors.

Because our \zmJ\ color criteria is slightly bluer than the $\zmJ >
0.9$ used by \cite{Bouwens2011a}, we effectively extend our redshift
selection window to somewhat lower redshifts.  Use of the
\cite{Bouwens2011a} color criteria, which was explicitly chosen to
exclude source with redshifts $z < 6.5$, would eliminate only one of
our candidates.

Using this criteria, we identified seven \zband-dropout galaxy
candidates with observed \Hband\ band magnitudes between $24.0-26.4$.
The candidates are named in decreasing order of their brightness in
the \Hband\ band, with the exception of the close pair A1703-zD5a and
A1703-zD5b.  As discussed in the next section, A1703-zD5a/5b most
likely represent two star-forming knots within a single source.  We
note that the A1703-zD5a component is detected in the \Vband\ band at
low significance ($2.1\sigma$), but not in the overlapping \gband\ and
\rband\ bands.  Thus, we interpret the slight \Vband-band detection as
a likely statistical fluctuation.

All of the candidates are clearly resolved with the exceptions of our
two faintest two candidates, A1703-zD6 and A1703-zD7.  A1703-zD6 is
unresolved with SExtractor stellarity parameter of 0.97.  A1703-zD7 is
somewhat extended, but only slightly resolved with a stellarity
parameter of 0.5.  Because A1703-zD6 is unresolved, under normal
circumstances it cannot be ruled out as a low-mass L,T dwarf star.
However, this candidate has subsequently been confirmed to be at
$z=7.045$ with deep Keck spectroscopic observations
\citep{Schenker2011}.

While the WFC3/IR data were taken after the ACS optical data, we can
rule we rule out supernovae as contaminants to our sample because the
sources are either resolved or in the case of the unresolved source
A1703-zD6, spectroscopically confirmed at $z=7.045$.

The positions of these sources in the Abell~1703 data are shown in
Figure~\ref{fig:a1703}.  Their properties are listed in
Table~\ref{tbl:photometry} and cutout stamps showing each of the
sources are presented in Figure~\ref{fig:stamps}.  The \zmJ\ and \JmH\
colors of the candidates are illustrated in Figure~\ref{fig:colcol}.
As seen in this figure, the \zmJ\ color of A1703-zD2 is exactly at our
selection limit (0.7).  While it could be in our $z\sim7$
\zband-dropout sample as a result of photometric scatter, this
candidate is completely undetected in the deep optical ACS data.  Its
somewhat bluer $\zmJ$ color means it is at the lower-redshift edge of
our \zband-dropout selection window.  This is completely consistent
with its photometric redshift of $z=6.4$ (see \S~\ref{sec:bpz}),
making it our lowest redshift candidate.

The best candidate, A1703-zD1, is an extremely bright \zband-dropout
candidate, with a \Hband\ magnitude of 24.0, that appears to be
resolved in three separate knots (see Fig.~\ref{fig:stamps} and
\S~\ref{sec:morph}).  In Figure~\ref{fig:hist} we present a histogram
of both the observed and intrinsic (unlensed) \Hband\ magnitudes
compared with the 73 $z\sim7$ candidates found in the HUDF09 and its
two parallel fields and the WFC3/IR Early Release Science observations
\citep{Bouwens2011a}.

\section{Source Magnifications and Counterimages}

Several detailed studies to model the lensing of Abell~1703 have been
performed in recent years \citep{Limousin2008, Richard2009,
Zitrin2010}.  We adopt the \cite{Zitrin2010} Abell~1703 strong lensing
model to estimate the magnifications of the seven $z\sim7$ sources and
to identify possible counterimages.  \cite{Zitrin2010} used 16
multiply-imaged systems behind Abell~1703 and applied two independent
strong lensing techniques to the high-quality, multiband ACS data,
yielding similar results.  Their strong lensing model places tight
constraints on the inner mass profile, and thus provides reliable
magnification estimates for background sources.  The magnifications of
the high-redshift candidates range from $\mu \sim 3$ to large
magnifications of $\sim25-40$, found for three of our sources that are
located near the critical curve, where the magnification formally
diverges.  The magnification of each candidate is presented in
Table~\ref{tbl:photometry}.

We estimated the magnification uncertainties by taking models
extracted from the $1\sigma$ confidence level, as determined by the
$\chi^2$ minimization of model, and marginalizing over the true
$1\sigma$ errors.  To make the error estimates more conservative, we
also incorporate the range of magnifications obtained within $\pm
0.5\arcsec$ of each candidate and apply a $\Delta z \pm 1.0$ to the
redshift of each source.  Thus, the $\pm 0.5\arcsec$ shift is a
measure of the magnification variance around the location of the
source and the application of $\Delta z \pm 1.0$ accounts for the
possible local uncertainty in the location of the critical curves.
For objects that are close to the critical lines, the magnification
errors are diverging due to their proximity to the critical curve,
while objects far away will have a well-determined magnification as
the latter slowly varies in regions away from the critical curve.

The brightest candidate, A1703-zD1, has a magnification of $\sim9$,
giving it an intrinsic magnitude of $\sim26.4$ in the \Hband\ band.
The Abell~1703 strong lensing model predicts counterimages for the
three high-magnification candidates, A1703-zD2 and the A1703-zD5a/5b
pair, which are located nearby or on the high-redshift critical curve
(see Fig.~\ref{fig:a1703}).  The lensing model predicts three
counterimages for A1703-zD2 ($\mu = 24.8$; see Fig.~\ref{fig:a1703}).
Taking into account the much smaller magnifications ($\mu = 5.5-9.0$)
of the counterimages, they are predicted to have \Hband\ magnitudes
between $26.0-26.5$.  This is sufficiently bright that there was some
possibility that we might locate them, but also a good chance we might
not because they could easily be lost in the wings of a foreground
galaxy.  Despite an extensive search, we did not find any viable
$z\sim7$ candidates near the predicted positions of the counterimages.

The close pair of A1703-zD5a and A1703-zD5b are also located in very
close proximity of the critical curve and as such have high
magnifications of $\mu\sim27-40$.  These candidates are also predicted
to have counterimages on the other side of the brightest cluster
galaxy (BCG; see Fig.~\ref{fig:a1703}) with magnifications of $\mu =
5.5$, about $5$ times less than A1703-zD5b.  The predicted
counterimages are expected to have an \Hband\ magnitude of $\ge27.0$,
which is fainter than our $5\sigma$ limiting magnitude of 26.9.
Hence, it is not surprising that no \zband-dropout candidates are
found in the predicted region of the counterimages.

The critical curve lies only $2.6\arcsec$ from passing between the
A1703-zD5a/b sources, which is within the model (and redshift)
uncertainty.  To test the hypothesis that these two candidates are
multiple images of the same source, we constructed a new model
assuming that zD5a and zD5b are the same object.  The resulting model
is physically plausible and the predicted counterimage of this system,
located in the same region marked in Fig.~\ref{fig:a1703}, is again
fainter than the $5\sigma$ limiting magnitude of 26.9.  Because the
overall reproduction of all other systems remains the same, we cannot
exclude this option based solely on the mass model.  This possiblity
could also be supported by their similar colors, morphologies, and
photometric redshifts.

However, because surface brightness is conserved in lensing, zD5a and
zD5b should have the same surface brightness if they are indeed the
same source.  For zD5a and zD5b, we find surface brightnesses of 25.9
and 25.8 mag arcsec$^{-2}$, respectively, in the \Jband\ band and 26.4
and 25.9 mag arcsec$^{-2}$, respectively in the \Hband\ band.  We note
that these results are consistent with \cite{Oesch2010b} who found a
mean \Jband-band observed surface brightness of $\sim26$ mag
arcsec$^{-2}$ for a sample of $z\sim7$ \zband-band dropouts candidates
spanning 26 to 29 mag in the \Jband\ band.  While their surface
brightness agrees in the \Jband\ band, zD5b has a brighter surface
brightness in the \Hband\ band.  On these grounds we conclude that
zD5a and zD5b are two unique sources.  Further, their very close
proximity of 0.76" in the image plane, with a magnification of 8.6
along the line separating them, translates to only $\sim480$ pc in the
source plane.  Thus, even though we do not observe a diffuse
component, which could have very low surface brightness below our
detection limits, between them, we conclude that zD5a and zD5b are
most likely two star-forming knots within a single source.  The
magnification of the combined A1703-zD5 source is
$31.9^{+20.3}_{-10.6}$.  We refer to the these sources separately
because they appear as distinct sources in our catalog.  The small
elliptical Kron apertures used to measure their colors are well
separated with sizes of $0.34\arcsec \times 0.26\arcsec$ and
$0.28\arcsec \times 0.22\arcsec$, respectively.

We also considered the possibilty that zD5a/b and zD2 were all
counterimages of the same source.  With zD5a/b representing two bright
star-forming knots in a single candidate galaxy and not counterimages
of the same source, we conclude that zD2 cannot be a counterimage of
the zD5a/b source based simply on morphology.  This was verified by
constructing an additional model considering the center of zD2 and the
center of zD5a/b as counter positions of the same source.

\ifemulateapj\fige\fi

\section{Photometric Redshifts}
\label{sec:bpz}

To estimate the redshifts of the candidates, we used the Bayesian
photometric redshift (BPZ) code \citep{Benitez2000, Benitez2004,
Coe2006}.  Briefly, the photometric redshifts are based on a $\chi^2$
fitting procedure to template spectra.  Because the shape of the
redshift distribution is not well calibrated at $z\sim7$, we assumed a
flat prior for all redshifts.  The photometric redshifts of the
\zband-dropout candidates are presented in Table~\ref{tbl:photometry}
and their posterior $P(z)$ probability distributions are shown in
Figure~\ref{fig:pz}.  We find redshifts in the range of $z_{phot} =
6.4-8.8$, with a median redshift of 6.7.

The posterior probability distributions show that there is a small
chance that A1703-zD3, A1703-zD5a, and A1703-zD7 could be at low
redshift.  Because of the modest $2.1\sigma$ detection in the F555W
band, the A1703-zD5a knot shows the highest probability of being at low
redshift.  While we cannot completely exclude the low-redshift
solutions for these candidates, objects in this magnitude range
($24.0-26.4$) with colors similar to LBGs would be rare and are more
likely to be at high redshift.  The BPZ results indicate that the
exceptionally bright candidate A1703-zD1 has a narrow probability
distribution at $z=6.7$ and has the highest probability of being at
high redshift, not showing any evidence to suggest a low-redshift
solution.

\ifemulateapj\tabb\fi

\section{Stellar Population Models}

We performed fits to the multiband \HST\ and \Spitzer\ photometry of
A1703-zD1 and A1703-zD4 using the stellar population models of
\cite{Bruzual2003}.  We adopted a \cite{Salpeter1955} initial mass
function (IMF) with mass cutoffs of 0.1 and 100 \Msun\ and models with
subsolar ($Z = 0.2 Z_{\sun} = 0.004$) metallicities.  The effect of
dust reddening is included in the models using the \cite{Calzetti2000}
obscuration law.  We use the \cite{Madau1995} procedure to correct the
models for Lyman-series line-blanketing and photoelectric absorption.
The stellar population models are constrained such that the stellar
age must be less than the age of the universe at the fit redshift
(e.g., 0.75 Gyr at $z = 7.0$).  We consider two star-formation
histories (SFH):  simple (single-burst) stellar population (SSP)
models and constant star-formation rate (CSFR) models.

The best-fit stellar population models for A1703-zD1 and A1703-zD4,
the two sources for which we were able to obtain IRAC
photometry, are shown in Figures~\ref{fig:zd1} and \ref{fig:zd4},
respectively and the parameters are given in Table~\ref{tbl:models}.
For these sources we find reasonably good model fits to the observed
broadband photometry.  For A1703-zD1, we note that the SED models are
unable to fit the low flux in the IRAC 4.5\micron\ band, resulting in
a somewhat higher $\chi^{2}_{\nu} = 1.3-2.4$

We find intrinsic (unlensed) stellar masses for both candidates in the
range from $(0.7-3.0) \times 10^{9}\ \Msun$ with star-formation rates
of $7.3 \pm 0.3$~\Msunyr and $8.2 \pm 1.2$~\Msunyr, broadly consistent
with those found for $z\sim6-8$ galaxy candidates \citep{Labbe2010a,
Labbe2010b, Gonzalez2010, McLure2011}.  Because these two candidates
are located near the edge of the WFC3/IR field, there are no source
templates for IRAC sources found beyond the WFC3/IR field of view.
This results in a large modeling error for the neighboring sources,
which we include in the total IRAC error for these candidates.  While
our modeling procedure weights each photometric data point by its
inverse variance, because of the relatively large IRAC errors for
these sources it should be noted that there is a larger uncertainty in
the resulting stellar masses and ages derived from the stellar
population models.

While we assumed a Salpeter IMF, fitting models using a
\cite{Chabrier2003} IMF would yield lower masses and star-formation
rates by a factor $\sim 1.5$.  For both candidates, we note the the
CSFR models provide consistently older SFR-weighted mean stellar ages
than that for the SSP models.  We find CSFR models with weighted ages
of $100-180$~Myr, values again similar to that recently reported for a
sample of $z\sim7-8$ candidates \citep{Labbe2010a, McLure2011}.

\ifemulateapj\figf\fi
\ifemulateapj\figg\fi

\section{Discussion}

\subsection{A1703-zD1 Brightness}

Some of the brightest ($\la26.5$) $z\ga5$ high-redshift galaxies to
date \citep{Franx1997, Frye2002, Kneib2004, Egami2005, Bradley2008,
Zheng2009, Bouwens2009a, Hall2011} have been identified in searches
behind strong lensing clusters.  Several of the more notable examples
include the bright \iband-dropout galaxy, A1703-iD1, found by
\cite{Zheng2009} behind Abell 1703.  This candidate has a NICMOS/NIC3
\Hband\ band magnitude of 23.9 and is lensed by a factor $\mu\sim3.1$.
From the SED fitting of this source, we found a photometric redshift
of $z=5.95 \pm 0.15$, which is consistent with the Keck spectroscopic
redshift of $z=5.827$ measured by \cite{Richard2009}.

At somewhat higher redshift, \cite{Kneib2004} identified an
exceptionally bright candidate (NIC3 \Hband = 24.1) behind the galaxy
cluster Abell~2218.  This triply-imaged candidate, A2218-iD1, lies
near the high-redshift critical curve and has a large magnification of
$\mu\sim25$.  While this very bright candidate has a similar
\Hband-band magnitude as A1703-zD1 and has been suggested to be at
$z\sim7$, it has a $\zband-J_{110}$ color of just $\sim$0.4
magnitudes, which is more consistent with a redshift of $z = 6.3 \pm
0.1$.  It therefore almost certainly has a lower redshift than the
present candidate A1703-zD1 and A1689-zD1, which is the brightest
$z\sim7.6$ candidate known.  A1689-zD1 was discovered by
\cite{Bradley2008} behind the massive galaxy cluster Abell~1689.  This
source is magnified by a factor of $\mu\sim9.3$ and has a NIC3 \Hband\
magnitude of 24.7, which is 0.7 magnitudes fainter than the
exceptionally bright candidate $z\sim7$ candidate A1703-zD1 presented
here.  Most recently, \cite{Hall2011} reported the discovery of a very
bright \zband-dropout candidate behind the Bullet Cluster.  The
lensing model of the Bullet Cluster suggests that this candidate is
doubly-imaged with observed \Hband-band magnitudes of 24.2 and 25.0
and magnifications of 8.4 and 12, respectively.  However, because this
candidate is detected in the \iband\ band, there is a fair chance this
object could be a low-redshift interloper \citep{Hall2011}.

\ifemulateapj\figh\fi

\subsection{A1703-zD1 Source-Plane Reconstruction and Morphology}
\label{sec:morph}

The large magnification of A1703-zD1 allows us an opportunity to
examine the morphology of this exceptionally bright $z\sim6.7$ galaxy
candidate at very high spatial resolution.  With a magnification of
$\mu\sim9$, the strong lensing effect provides an increased spatial
resolution by about a factor $\sim3$ compared to an unlensed source.
This permits us to resolve spatial structures that would otherwise be
unobservable in high-redshift $z\sim7$ galaxies.

We used the \cite{Zitrin2010} Abell~1703 cluster lensing model to
reconstruct A1703-zD1 in the source plane at $z\sim6.7$.  The
deprojected image of A1703-zD1 in the WFC3/IR \Jband\ band is shown in
Figure~\ref{fig:zd1src}.  The linear magnification along the shear
direction is $4.88^{+0.11}_{-0.14}$ and $1.84 \pm 0.01$ in the
direction perpendicular to the shear direction.  This candidate
appears to be comprised of three resolved star-forming knots, each
with a radius $r\sim0.08\arcsec$ ($0.4$~kpc) in the source plane.
Altogether, A1703-zD1 has an extended linear morphology that spans
$\sim0.74\arcsec$ ($\sim4$~kpc) in the source plane at $z=6.7$.  Of
course the physical size and structure we infer for this candidate is
somewhat dependent on the details of the gravitational lensing model.

There are now several examples of $z\ga5$ lensed galaxy candidates
that have morphologies consisting of star-forming knots
\citep{Franx1997, Kneib2004, Bradley2008, Zheng2009, Swinbank2009,
Zitrin2011}.  Interestingly, each of the bright lensed candidates
discussed in the previous section are extended and show significant
substructure.  In particular, both A1703-iD1 and A1689-zD1 show a pair
of resolved star-forming knots.  Additionally the pair of $z\sim6.5$
dropout candidates, CL0024-iD1 and CL0024-zD1, each seem to consist of
two components.  With a separation of only 2~kpc in the source plane
and nearly identical redshifts and properties \citep{Zheng2009}, it is
possible that CL0024-iD1 and CL0024-zD1 are spatially associated or
merging galaxies.  Recently, \cite{Oesch2010b} even found extended
features with resolved double cores in two unlensed $z\sim7$ galaxies
identified in the WFC3/IR HUDF.  The apparent frequency of
high-redshift galaxies showing substructure and multiple components is
perhaps not surprising, but it provides strong evidence that both
clumpy star formation and merging are important aspects of galaxy
buildup at these very early epochs in the universe.  The existence of
substantial substructure is also expected based on studies of
low-redshift Lyman-break analog galaxies \citep{Overzier2008}.

\subsection{Number Counts of $z\sim7$ Candidates}

The discovery of seven $z\sim7$ \zband-dropout candidates in a single
cluster field is quite remarkable, especially since four of our
candidates (zD1, zD3, zD6, and zD7) lie in a small region near the
edge of the WFC3/IR image.  While this may be surprising and call into
question the reliability of these candidates, our confidence is
bolstered by the spectroscopic confirmation at $z=7.045$
\citep{Schenker2011} of A1703-zD6, the second-faintest candidate.  We
consider the extremely bright (\Hband\ = 24.0) candidate A1703-zD1,
with a strong $\zmJ$ break of 1.7, to be a rather robust high-redshift
candidate, but we acknowledge that A1703-zD7, the faintest and nearly
unresolved candidate, is certainly less secure than the other
candidates.

While the cluster magnification increases the effective depth of the
observations, the source-plane area that is surveyed at high redshift
decreases inversely proportional to the magnification factor.  The
Abell~1703 WFC3/IR image field of view is 4.6 arcmin$^{2}$, but we
estimate that we are effectively surveying only $\sim0.9$ arcmin$^{2}$
in the $z\sim7$ source plane.  Thus, we derive a simple estimate of
the number density of $z\sim7$ sources in Abell~1703 of 7.8
arcmin$^{-2}$.  Typically, blank field surveys such as the HUDF09 and
its two parallel fields \citep{Bouwens2011b, Oesch2010a} find $\sim
3.5-4.3$ $z\sim7$ sources per arcmin$^{2}$.  While our number density
is larger than that found in typical blank fields, if we allow for the
possibility that zD7 is a low-redshift interloper, we are left with
six sources or 6.7 arcmin$^{-2}$.  However, we note that cosmic
variance in the number counts of high-redshift sources is significant
in these relatively small-area \HST\ fields \citep{Trenti2008},
especially the cluster fields with their significantly reduced area in
the high-redshift source plane.

We can place the observations of Abell~1703 in context with those
obtained for a small, but growing, sample of strong lensing clusters
with high-quality optical and NIR multiband data.  While
\cite{Hall2011} has reported the discovery of ten \zband-dropouts
behind the Bullet Cluster \citep{Hall2011}, most cluster lensing
fields have produced at most $1-2$ $z\ga7$ candidates.  This includes
NICMOS and WFC3/IR imaging of Abell~1689 and the recent 16-band
imaging of Abell~383 and MACS1149 obtained by the CLASH MCT program.
The apparent large variations in the number of $z\ga7$ candidates
discovered behind lensing clusters suggests, not surprisingly, that
$z\ga7$ galaxies may be highly clustered.  One may therefore need to
survey a large number of clusters to overcome the substantial
large-scale structure effects.

\section{Summary}

We report the discovery of a very bright, highly magnified LBG
candidate (A1703-zD1) at $z\sim6.7$ behind the massive galaxy cluster
Abell~1703.  A1703-zD1 is 0.2 magnitudes brighter than the recently
discovered \zband-dropout candidate behind the Bullet Cluster
\citep{Hall2011} and 0.7 magnitudes brighter than the current
brightest known $z\sim7.6$ galaxy A1689-zD1, identified behind the
massive galaxy cluster Abell~1689 \citep{Bradley2008}.  We find a
strong \zmJ\ break of at least 1.7 mag and best-fit photometric
redshift of $z=6.7$.  Using the \cite{Zitrin2010} cluster lensing
model, we estimate a magnification of $\mu = 9$ at $z\sim6.7$ at the
position of A1703-zD1.  The candidate is extended, spanning
$\sim4$~kpc in the reconstructed source plane, and is resolved into
three resolved star-forming knots.  The source plane deprojection
shows that the star formation is occurring in compact knots of size
$\sim0.4$~kpc.

Additionally, we find six other bright $z\sim7$ \zband-dropout
galaxy candidates behind Abell~1703.  One of these candidates,
A1703-zD6, has been subsequently confirmed with Keck spectroscopy to be
at $z=7.045$ \cite{Schenker2011}.  The candidates are observed with
\Hband\ band magnitudes of $24.9-26.4$, with a wide range of
magnifications from $3-40$.  Their photometric redshifts are found to
be in the the range of $z_{phot} = 6.4-8.8$, with a median redshift of
6.7.  Using stellar population models to fit the rest-frame UV and
optical fluxes for A1703-zD1 and A1703-zD4, we derive best-fit values
for stellar masses $(0.7 - 3.0) \times 10^{9}\ \Msun$, stellar ages $5
- 180$~Myr, and star-formation rates $\sim7.8$~\Msunyr.

A1703-zD1, with a photometric redshift of $z\sim6.7$ is the brightest
observed $z\sim7$ galaxy candidate found to date.  We are planning to
observe A1703-zD1 with near-IR spectroscopy to confirm its redshift
and study its spectrum.  Bright high-redshift galaxies such as these
are valuable targets for ground-based spectroscopy and are pathfinding
objects for future facilities such as the {\em James Webb Space
Telescope}.

\acknowledgments

We would like to thank Ivo Labb\'{e} and Valentino Gonz\'{a}lez for
assistance with the \Spitzer/IRAC photometry.  We also thank the
anonymous referee whose comments and suggestions significantly
improved the quality and clarity of this work.  ACS was developed
under NASA contract NAS 5-32865, and this research has been supported
by NASA grants NAG5-7697 and HST-GO10150.01-A and by an equipment
grant from Sun Microsystems, Inc.  The {Space Telescope Science
Institute} is operated by AURA Inc., under NASA contract NAS5-26555.
Archival data were obtained from observations made by the {\em Spitzer
Space Telescope}, which is operated by the Jet Propulsion Laboratory,
California Institute of Technology under NASA contract 1407.


\ifemulateapj\else
    \taba
    \clearpage
    \tabb
    \clearpage
    \figa
    \clearpage
    \figb
    \clearpage
    \figc
    \clearpage
    \figd
    \clearpage
    \fige
    \clearpage
    \figf
    \clearpage
    \figg
    \clearpage
    \figh
\fi

\end{document}